\documentclass[12pt]{iopart}
\usepackage{graphicx}
\usepackage{psfrag}
\begin{document}

\title{Dynamics of  random dipoles  : chaos {\it vs} 
ferromagnetism}
\author{F.~Borgonovi}
\address{Dipartimento di Matematica e Fisica, Universit\`a Cattolica, via Musei 41, 25121
Brescia, Italy}
\address{I.N.F.N., Sezione di Pavia, Via Bassi 6, 27100, Pavia, Italy}
\author{G.~L.~Celardo}
\address{Dipartimento di Matematica e Fisica, Universit\`a Cattolica, via Musei 41, 25121, Brescia, Italy}
\address{I.N.F.N., Sezione di Pavia, Via Bassi 6, 27100, Pavia, Italy}
\begin{abstract}
The microcanonical dynamics of an ensemble of random magnetic dipoles 
in a needle has been investigated. 
Due to the presence of a constant of motion in the 1--D case,
a ``dimensional'' phase transition in the quasi one dimensional case has been 
found separating a paramagnetic chaotic phase from a ferromagnetic
regular one. 
In particular, a simple criterium for the 
transition has been formulated and an intensive critical parameter found. 
Numerical simulations support our understanding of this complex phenomenon.
\end{abstract}
\date{\today}
\maketitle

\section{Introduction}

A truly comprehensive understanding of magnetism at the nanoscale 
is still lacking and has important consequences 
in the technology of memory  and information processing  devices.

Many unsolved problems about magnetic properties of 
diluted spin systems attracted recently great attention.
Among the open problems there is the
emergence of ferromagnetism in doped diluted 
systems\cite{sangaletti}, where the Curie
temperatures can be as high as $300$ K, and a deep
theoretical understanding of the 
magnetic properties of dilute dipole systems (spin glass transition,
ferromagnetic and anti-ferromagnetic transitions).   

Here we will concentrate on randomly arranged dilute classical dipoles,
which are called dipole glasses. Many results in literature,  
sometimes controversial, exist on such kind of systems.
Magnetic properties of dipole-dipole interacting spins are 
particularly difficult to study due to many factors:
long range nature of the interaction, anisotropy and frustration. 
Long range and anisotropy can induce ergodicity breaking\cite{palmer}
in a system.
Breaking of ergodicity, a concept introduced by Palmer\cite{palmer}, 
and recently found explicitly \cite{jsp,TNT} in a class of long-ranged 
anisotropic spin systems, 
is a key word to understand phase transitions too, even 
if it should not be confused with breaking of symmetry\cite{paperergpt}.
Speaking loosely, few constants of motion, 
such as the energy, or the angular momentum,
in a particular geometry,
produce a separation of the allowable phase space 
in two or more subspace over which the motion  is constrained.
In Ref.\cite{jsp} the energy at which the separation occurs has been calculated
explicitly  for an anisotropic 1--D classical Heisenberg systems. 
In that case both 
the anisotropy and the long ranged nature\cite{libroruffo} 
of the inter-spin interaction,
are  essential ingredients in order to have breaking of ergodicity\cite{vachal}.
On the other hand, frustration, that is the impossibility to
attain a global minimal energy minimizing locally the interactions,
induces a dependence of the ferromagnetic and anti-ferromagnetic properties 
on the lattice geometry \cite{zhang12}.

Other  results  concerning the so-called Ising dipole glass can also
be found in literature, where 
Ising simply means  uni-axial. To quote but a few: 
spin glass transition for high concentration,
using Monte Carlo simulation\cite{snider7,snider21}, 
mean field spin glass transition
at low concentration depending on the lattice geometry\cite{snider6},
absence of spin glass transition for low concentration using Wang-Landau 
Monte Carlo simulations
\cite{snider} or the recent spin glass transition at non zero temperature from
extensive numerical simulation\cite{condmat08100854v2}.

In this paper  we will focus our analysis on a
dipole glass of freely rotating classical dipoles.
First of all the dipole glass is a typical example of very frustrated system
\cite{leshouches, vug, zhangwidom}, so that different ground
state configurations can exist depending on the geometry and the spin
concentration.
Results in the canonical ensemble typically consider a mean field approach,
and it is common lore that the random
positions of the dipoles induce  magnetic field fluctuations.
These fluctuations do not vanish   at $T\to 0$, unlike thermal fluctuations,
and tend to suppress magnetic order even at $T=0$\cite{vug, zhangwidom}.
So, magnetic order,
is expected to happen only for high impurity concentration 
(and small temperature) \cite{zhangwidom,snider5,snider20}.
Mean field theories consider only the equilibrium properties
and do not take into account the time needed to reach the equilibrium
situation and finite size effects. 
On the other hand the question of how long
a metastable state can last is a major issue in determining 
the magnetic properties of a system.

In this paper we study  the microcanonical
dynamics, reserving the study of the influence of a thermal bath
for further investigations.
We analyze the microcanonical dynamics of dipoles put 
at the vertexes of a cubic lattice 
(so that their relative distance cannot be smaller
than the lattice size), only on the basis of the Landau-Lifshitz-Gilbert
equations of motion.
3--D dipole-dipole interacting systems can be realized quite easily
in laboratory, for instance doping a non magnetic media  with paramagnetic ions,
weakly interacting with the lattice and with an inter-dipole distance 
sufficiently large in order to neglect Heisenberg interaction (
low concentration).
The choice of studying  a random glass instead 
of a system composed of dipoles regularly arranged
in some lattice is twofold: from one hand is it relatively easy to dope a system
putting some paramagnetic doping ions in a random way inside 
any non-magnetic media.
On the other hand, a  3-D cubic lattice with a full concentration of dopant 
ions $\delta = 1$, even
if thin, does not  have a ferromagnetic 
ground state \cite{sauer}, so that another
type of transition should be considered (paramagnetic/antiferromagnetic) .

Anticipating some of the results, we have found that taking into 
account a typical
experimental situation with needle-shaped sample, 
at low dipole concentration a further constant of motion
appears that induce a kind of ``phase'' transition related
to invariant tori, which separate 
the allowable phase space in many disconnected regions. 
This result seems to indicate that at very low concentration a 
system of random dipoles in a needle resemble a one dimensional arrangement 
of dipoles, and thus can have ferromagnetic behavior.
In this particular case, the ergodicity breaking is not due to an increase
of energy, but to an increase of perturbation, which means the tendency
to a transformation from a needle shape (quasi 1--D system) to a cubic shape
(3--D shape).
In a sense, these results are more akin to the standard perturbation
theory in classical dynamical systems\cite{chirikov, lichtenberg},
re-interpreted in the light of phase transitions
induced by  demagnetization times \cite{firenze}. 

In the future we are going  to study the same system in contact 
with a thermal bath.
In this case the presence of the ergodicity breaking found in  \cite{jsp,TNT}
should influence the demagnetization times. 
In the microcanonical case, the
presence of this ergodicity breaking is hidden by the quasi-integrability
of motion.

\section{The Classical Model and the Perturbative Approach}

\begin{figure}[h!]
\begin{center}
\includegraphics[scale=0.5]{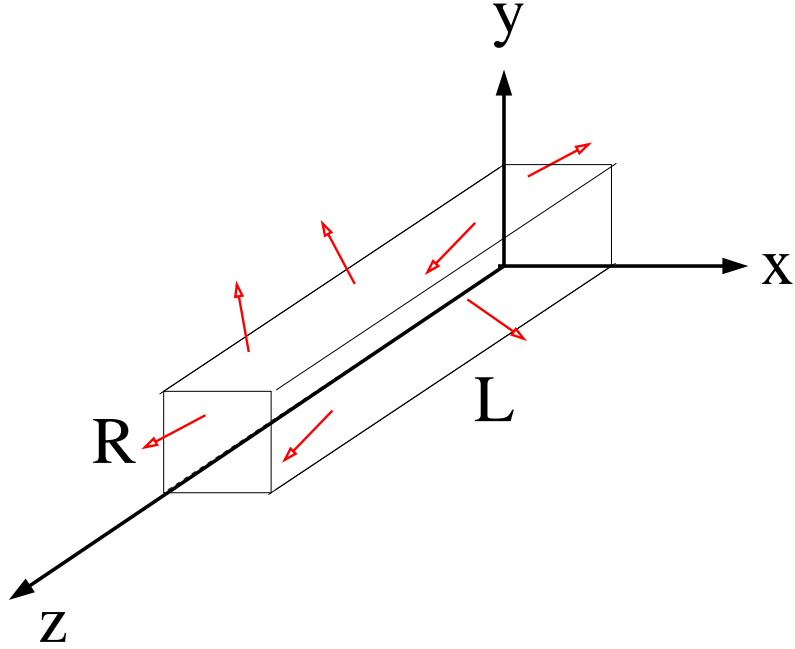}
\caption{Needle geometry. The classical dipoles are put in a random 
way on the vertexes of a cubic lattice of size $a$. $R$ and $L$ are given
in units of the lattice size $a$.}
\label{zero}
\end{center}
\end{figure}

Let us consider a system of $N$ classical dipoles $\vec{\mu}_i$ randomly put
at the nodes of  a 3--D gridded box $R \times R \times L$, with $L \gg R$, and low
concentration $\delta \ll  1$, as indicated in  Fig.~\ref{zero},

\noindent
From the physical point of view  it represents
a dilute system of paramagnetic
ions in a non magnetic bulk, 
with  a concentration   $\delta = N/N_s$ where $N_s= R^2 L$ is the number
of allowable sites in the 3--D lattice.
As explained above, such a system  can be realized in 
laboratory, doping a non--magnetic system having a cubic lattice
with  paramagnetic impurities
(e.g. doped $Ti O_2$ and other\cite{sangaletti}).
If the  dipoles weakly interact with the lattice and if their
average distance is much greater than the Bohr radius, we can simply neglect 
the Heisenberg (exchange) interaction and represent 
their mutual interaction and 
dynamics with a pure dipole-dipole interaction energy:

\begin{equation}
E = \frac{\mu_0 \mu^2 }{4\pi a^3 }  \sum_{i=1}^N \sum_{j> i} 
\frac{1}{|r_{ij}|^3} \left[ \vec{S}_i \cdot \vec{S}_j -
3(\vec{S}_i \cdot \hat{r}_{ij}) (\vec{S}_j \cdot \hat{r}_{ij} )
\right].
\label{en}
\end{equation}
Here $\vec{S}_i$ is the $i$--th dimensionless spin vector 
\begin{equation}
\vec{S}_i \cdot \vec{S}_i =1,
\label{conm}
\end{equation}
 $\mu$ is the magnetic moment of the paramagnetic  doping ions
and $r_{ij}$ is the distance between the i-th and the j-th spin in
units of the lattice spacing $a$.

The dynamics is described  by the Landau-Lifshitz-Gilbert equations
of motion:
\begin{equation}
\frac{d}{dt} \vec{\mu}_k = \gamma \vec{\mu}_k \times \frac{\delta E}
{\delta \vec{\mu}_k },
\label{LLG}
\end{equation}
where $\vec{\mu}_k = \mu \vec{S}_k$ and $\gamma$ is the gyromagnetic ratio.

\noindent
They can be rewritten in the dimensionless form, 

\begin{equation}
\frac{d}{d\tau} \vec{S}_k = \vec{S}_k \times \frac{\delta E_0}
{\delta \vec{S}_k },
\label{LLGd}
\end{equation}
where the following dimensionless quantities have been introduced:
\begin{equation}
\begin{array}{lll}
E_0 &= E\displaystyle  \frac{4\pi a^3}{\mu_0\mu^2}\\
&\\
\tau &= \omega t, \quad {\rm with} \quad \omega = \displaystyle
\frac{\gamma \mu \mu_0}{4\pi a^3}.\\
&
\label{diml}
\end{array}
\end{equation}

The system of equations  (\ref{LLGd})
conserves the energy (\ref{en})   and the squared modulii of the spins 
(\ref{conm}). 

The diluted  quasi 1--D system  can be  magnetized with a 
strong magnetic field directed along $L$, the longest axis 
($z$-axis).
The questions we would like to answer is the following: 
What is the dependence of the average  demagnetization time
and its fluctuations on the system parameters? 

The relevant parameters to take into account are  the concentration $\delta$ of 
paramagnetic ions and  the aspect ratio $\epsilon = R/L$. In principle,  due
to the long-ranged nature of the dipole interaction, one could ask
whether  there are effects dependent on both the system size
and the number of doping spins $N$, even if in quasi 1--D systems, the
dipole interaction can be treated as  a short range interaction.

From the point of view of the equations of motion (\ref{LLGd}), 
if the $N$ dipoles are lying along a straight line 
($R=0 \Rightarrow \epsilon=0)$, there is a further constant of motion,
i.e. $M_z = (1/N) \sum_k S_k^z$. Therefore, for a 1--D system,
 the answer to the first question above
is very simple :
a state with any initial magnetization $M_z(0) \ne 0 $ will keep 
the initial magnetization forever.  
The natural question thus becomes: what happens
for $\epsilon \ne 0$? Will a magnetized state  demagnetize and how much 
time it takes to do that?

\begin{figure}[t]
\begin{center}
\vskip 1truecm
\includegraphics[scale=0.4,angle=0]{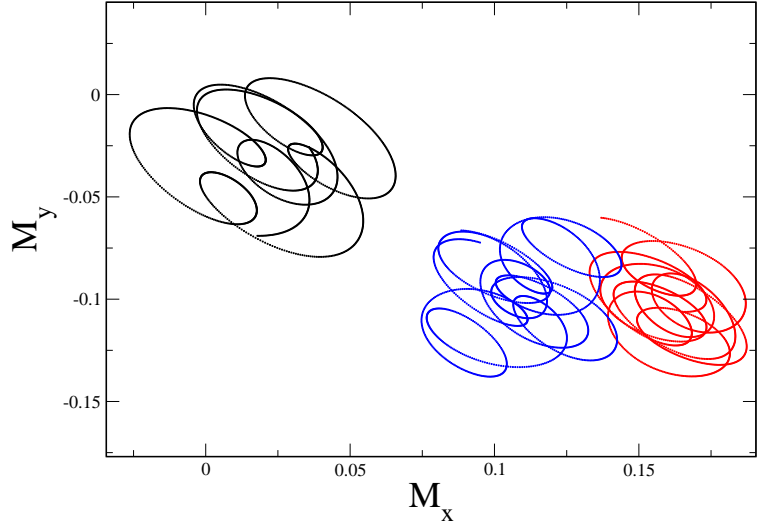}
\caption{Three different trajectories of the total magnetization
at the plane $M_z=0$, for $R=4$, $L=4000$, 
$\delta=10^{-3}$, $N=64$, in the integrable case.
Initially spins are chosen with random components on the unit sphere.
}
\label{int}
\end{center}
\end{figure}

The classical dynamical picture can be simplified adopting 
a perturbative approach, namely
approximating the unit versor between two spins as follows:
\begin{equation}
\label{app1}
\hat{r}_{ij} =
\cos \theta_{ij} \hat{z} + \sin \theta_{ij} ( \cos \phi_{ij} \hat{x} +
\sin \phi_{ij} \hat{y} ) \simeq 
\hat{z} + (\epsilon N ) ( \cos \phi_{ij} \hat{x} +
\sin \phi_{ij} \hat{y} ) 
\end{equation}
where $\hat{x}, \hat{y}, \hat{z}$ are the unit versors, $\phi_{ij}$ are
the azimuthal
angles with respect the $z$-axis and $\theta_{i,j}$ are the polar
angles. In the last equation we approximate $\cos \theta_{ij} \simeq 1$
and $\sin \theta_{i,j} \simeq  R/\langle d \rangle$, where, for dilute
dipoles in a needle goemetry, 
$\langle d \rangle \simeq L/N$ is the average distance among spins.
The energy (\ref{en}), 
to first order in $\epsilon N $, becomes:
$E_0 = H_0 + \epsilon N V$, 
where $H_0$ is the energy part that conserve $M_z$,
and $V$ is the perturbation,
\begin{equation}
\begin{array}{lll}
\label{ham11}
H_0 &=& \displaystyle
\frac{1 }{2 }  \sum_{i=1}^N \sum_{j\ne i}
\frac{1}{|r_{ij}|^3} \left[ S_i^x S_j^x + S_i^y S_j^y  -
2 S_i^z S_j^z,
\right]\\
V &=&
\displaystyle -3  \sum_{i=1}^N \sum_{j\ne i}
\frac{1}{|r_{ij}|^3} \left[ \cos \phi_{ij} S_i^z S_j^x + 
\sin \phi_{ij} S_i^z S_j^y  
\right].
\end{array}
\end{equation}
The equations of motion for the macroscopic variables, 
$M_{x,y,z}$ 
can be written as,
\begin{eqnarray}
\displaystyle \frac{dM_z}{d\tau} &=&  3\epsilon  \sum_{i\ne k} 
\frac{1}{|r_{ik}|^3}
S_i^z \left( S_k^y \cos \phi_{ik} - S_k^x \sin \phi_{ik} \right)\\
&\nonumber \\ 
\displaystyle \frac{dM_y}{d\tau} &=&  \frac{3}{N}  \sum_{i\ne k} 
\frac{1}{|r_{ik}|^3}
\left\{ S_i^z S_k^x + 
\epsilon N \left[
S_k^x  S_i^y \sin \phi_{ik} +( S_k^x S_i^x - S_k^z S_i^z )\cos \phi_{ik} 
\right] \right\} \nonumber \\
&\nonumber \\
\displaystyle \frac{dM_x}{d\tau} &=&  - \frac{3}{N} \sum_{i\ne k} 
\frac{1}{|r_{ik}|^3}
\{ S_i^z S_k^y +
\epsilon N \left[
S_k^x  S_i^y \cos \phi_{ik} +( S_k^y S_i^y - 
S_k^z S_i^z )\sin \phi_{ik} \right] \}, \nonumber \\ \nonumber
\label{eqmot}
\end{eqnarray}
and, in particular, for $
\epsilon=0$ we have :

\begin{equation}
\begin{array}{lll}
\displaystyle \frac{dM_z}{d\tau} &=  0\\
&\\
\displaystyle\frac{dM_y}{d\tau} &=  \displaystyle\frac{1}{N} \sum_k \omega_k  S_k^x \\
&\\
\displaystyle\frac{dM_x}{d\tau} &=  \displaystyle -\frac{1}{N} \sum_k \omega_k  S_k^y,
\end{array}
\label{eqmot0}
\end{equation}
having defined, the average ``local'' frequencies:
\begin{equation}
\omega_k = 3 \sum_{i\ne k } \frac{1}{|r_{ik}|^3} S_i^z.
\label{freq1}
\end{equation}
These equations describe a kind of rotation in the plane perpendicular 
to the  $z$--magnetization (which is a constant of motion).
Therefore one could expect that for $\epsilon N  \ll 1 $ a  rotational-like motion
about the $z$--axis persists, while $M_z$ remains a quasi constant of motion.
This is what can be observed for instance by a direct inspection of the trajectories
of the macroscopic vector $\vec{M}$, in the plane $x,y$,
see Fig.~\ref{int},  where few selected trajectories has been iterated in time,
for $\epsilon= 10^{-3}$ and $N=64$.
Quite naturally, on increasing the perturbation strength $\epsilon$, one 
could expect that  the invariant tori $M_z = const$ will be broken,
and, eventually,  a stochastic motion of the 
macroscopic variable $M_z$ will appear.
In the next Section we will study the survival of invariant tori, 
under the dimensional perturbation $\epsilon N > 0 $.

\section{The chaotic--paramagnetic  and the integrable--ferromagnetic phases}

The dynamical behavior of
the system can be characterized by a ``regular region'' 
$\epsilon N < 1$  in which the 
magnetization
$M_z (\tau)$ is bounded in a small interval $\delta M_z$, while,
for $\epsilon N > 1$,  $M_z (\tau)$  quickly decays to zero.
To be more precise, the transition across $(\epsilon N)_{cr} \simeq 1$ 
is smooth, namely
there is a region of $\epsilon N$ values in which 
the initial magnetization decay to some non zero constant
when the time $\tau$ becomes large.

\begin{figure}[t]
\begin{center}
\vskip 1truecm
\includegraphics[scale=0.4,angle=0]{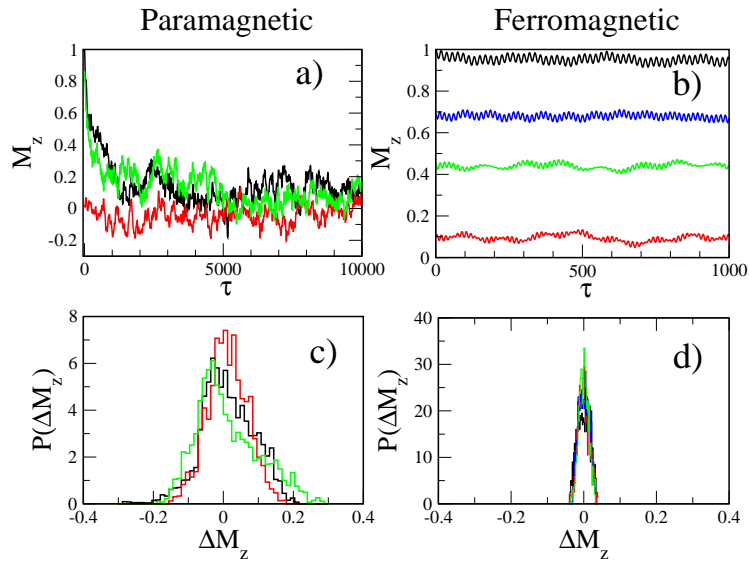}
\caption{Data in this Figure refer to systems with $N=64$ spins and a 
concentration $\delta = 10^{-3}$.
The time behavior of the magnetization is shown, for different initial conditions,
in the overcritical case $\epsilon = 0.125$ (a) ($L=160, R=20$) and in the
undercritical case $\epsilon = 10^{-3}$ (b)  ($L=4000, R=4$).
 In c) and d) the probability
distribution functions for
 the fluctuations
$\Delta M_z = (\langle M_z^2 \rangle-\langle M_z\rangle^2)^{1/2}$
around the equilibrium value is shown
for the data given respectively in a) and b).
}
\label{uno}
\end{center}
\end{figure}

The critical value of the perturbation strength $(\epsilon N)_{cr} \simeq 
1$ can be 
obtained with the following
hand-waving argument. Let us divide the 3--D box in $n=L/R=1/\epsilon$ 
small cubic boxes of side $R$. If the impurities concentration  $\delta$
is sufficiently small in order to have only one spin inside
each $R$-side box then the system is approximately one dimensional
and $M_z$ can be considered an approximate constant of motion.
Otherwise, for large $\delta$, the system behaves like a 3--D system 
and $M_z$ can spread everywhere. 
In other words, in order to have less than one spin in each $R^3$ block
one should have $N/n < 1$, or $\epsilon N  < (\epsilon N)_{cr} = 1$.

Moreover the study of the dynamics done in the previous Section
suggests to take as a small parameter $\epsilon N $ and to look for
ferromagnetism when $\epsilon N < 1$. Note that this choice is also
appropriate from the thermodynamic point of view since 
$\epsilon N = R N/L $ is an intensive parameter in the 
large $N$ limit $L \to \infty, \quad N\to \infty, \quad N/L = const$,
with $R$ fixed.

An example is shown in Fig.~\ref{uno}, where the dynamics of magnetization
has been plot in the overcritical case ($\epsilon N > 1$ 
Fig.~\ref{uno}a) and in the undercritical one 
($\epsilon N< 1$
Fig.~\ref{uno}b). Different trajectories, corresponding to 
different  initial conditions $M_z(0)$ have been shown in different colors.
As one can see, in the ``paramagnetic'' phase ($\epsilon N > 1$)
the magnetization first decays to zero and then it fluctuates 
randomly around zero.
On the contrary, in the ``ferromagnetic'' phase ($\epsilon N< 1 $),
it shows a periodic behavior around the  initial conditions.

This behavior is quite typical in the study of dynamical systems, where
the increase of a suitable perturbative parameter
 is related to the breaking of invariant tori and to the emergence
of chaotic motion
\cite{chirikov,lichtenberg}.

It is also remarkable to study the  fluctuations around 
the  asymptotic behavior:
in the undercritical case (Fig.~\ref{uno}d) fluctuations 
are much smaller than  in the 
overcritical case (Fig.~\ref{uno}c), roughly 10 times for this case, 
as can be seen comparing the width of the probability distribution functions
in Fig.~\ref{uno} c) and d).

\begin{figure}[t]
\begin{center}
\vskip 1truecm
\includegraphics[scale=0.4, angle=0]{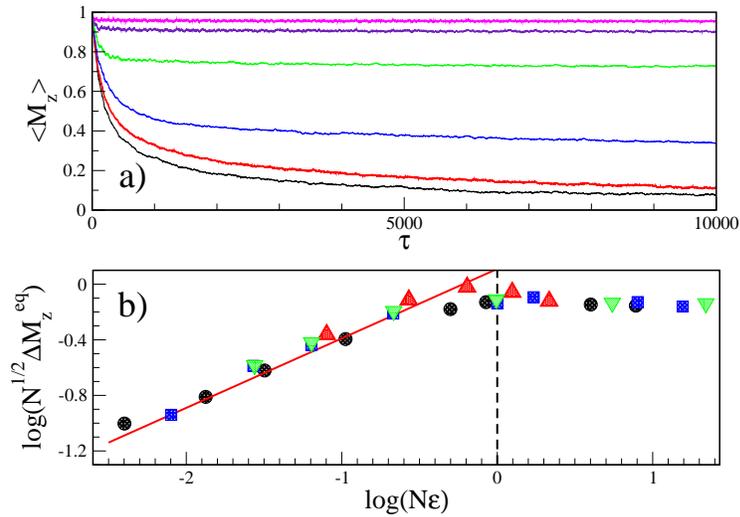}
\caption{a) Time behavior of the average magnetization for different
values of the aspect ratio $\epsilon = 1.25 \times 10^{-4},
2.9 \times 10^{-4}, 9.8 \times 10^{-4},
4.5 \times 10^{-3}, 2.5 \times 10^{-2}, 10^{-1} $ (from the upper to the lower),
fixed number of spins $N=220$ and fixed concentration $\delta= 10^{-3}$.
The average is taken over $100$ different random configurations.
Initially we choose $S_i^z(0)=1$, $i=1,\ldots,N$.
b) Dependence of the equilibrium value of fluctuations as a function
of $\epsilon N $. Dashed vertical line indicates the critical value
$\epsilon  N  = 1$.
Red line indicates the dependence $\sqrt{N \epsilon}$.
Different symbols stand for : $N=40, \  \delta = 0.5\ 10^{-3}$ (circles),
$N=80,\ \delta = 10^{-3}$ (squares), $N=220,\  \delta = 10^{-3}$ (triangles down),
$\epsilon=0.1, \  \delta = 10^{-2}$ (triangles up).
}
\label{due}
\end{center}
\end{figure}

The large fluctuations around the average values and in order 
to fit a possible experimental situation,
suggest  averaging  over disorder, namely
an ensemble of samples  with different random
configurations, initially magnetized along the $z$--axis.

The results for the ensemble average $\langle M_z \rangle$
are shown in Fig.\ref{due}a, where the different behavior 
in 
the two ``phases''  $\epsilon N  <  1 $  and
$\epsilon N  > 1 $ are reflected in an average magnetization
not decreasing or decreasing to zero. Indeed, the average magnetization 
in the undercritical regime reaches some equilibrium value
different from zero after some initial decay, while 
in the overcritical regime it goes to zero in an algebraic way.

Ensemble fluctuations at the equilibrium are 
independent of $\epsilon N $ in the paramagnetic phase
while  in the ferromagnetic one  are typically smaller and increasing as
$\sqrt{\epsilon N }$.
They are presented in Fig.~\ref{due}b), where
$$\Delta M_z^{eq} = \lim_{\tau\to\infty} \Delta M_z(\tau)$$
has been shown as a function of $\epsilon N $.
On the vertical axis we renormalize the asymptotic values by $\sqrt(N)$,
to take into account fluctuations due to variation of the number of spins $N$.
Each set  of points on the plot corresponds to an ensemble 
of magnetized needles,
with the same concentration $\delta$ and number of spins $N$  (paramagnetic ions) and 
different aspect ratio $\epsilon$, or 
same concentration and aspect ratio and different number of  spins.
It is quite remarkable that the critical
value $\epsilon N \sim 1 $ , is well fitted  by all different series, suggesting
$\epsilon N $ as a good scaling parameter for the macroscopic behavior.

Both the independence of the perturbation strength in the paramagnetic
phase and the square root
dependence on $\epsilon N $ in the ferromagnetic phase
 can be understood on the basis of classical dynamical theory. 
Breaking invariant tori with a perturbation strength $  k $ corresponds
to create stochastic layers between invariant tori whose size is
proportional to $\sqrt{k} $\cite{chirikov,lichtenberg}.
On the other side when the system is completely chaotic, since the variable 
$M_z$ is bounded, it can only occupy all the allowable stochastic region,
and a further increasing of perturbation strength can not modify this size.

Finally we point out that around $\epsilon N=1$ we can expect a transition
from a ferromagnetic ground state to an antiferromagnetic ground state.
Indeed for $\epsilon N \ll 1$ the system is close to a $1D$ 
arrangement of dipoles, so that the ground state will be
ferromagnetic as pointed out in the previous Section. 
On the other side for $\epsilon N \gg 1$
the system is close to a $3D$ arrangement of dipoles.
In this case, for a simple cubic lattice, 
it can be shown \cite{sauer} 
that the ground state is antiferromagnetic.
Some numerical simulations we did confirm this conjecture,
but work is still in progress and will be presented in a future publication.

\section{Demagnetization time}

\begin{figure}[t]
\begin{center}
\vskip 1truecm
\includegraphics[scale=0.4, angle=0]{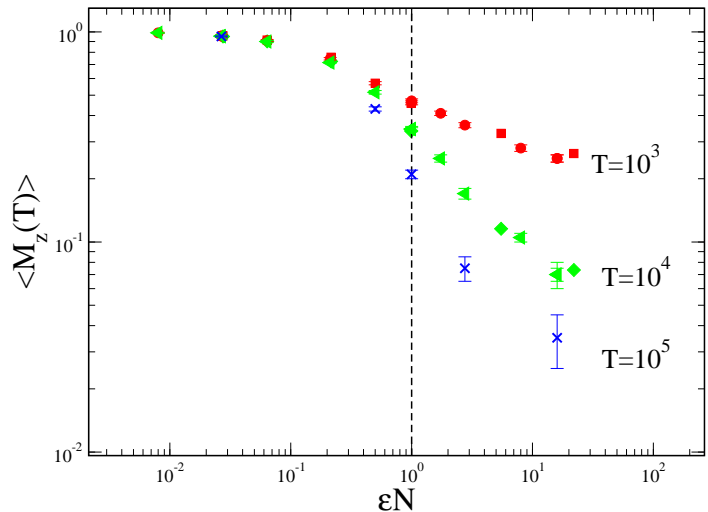}
\caption{Average Magnetization at time $T$ as a function of $\epsilon N$. 
For the same  $\epsilon N$ we show the average magnetization after 
different simulation times, $T$. Data refer to the case $\delta=10^{-3}$. 
Different number of dipoles  are shown:  
$N=220$ (circles for $T=10^3$ and
lozanges for $T=10^4$) and $N=80$ 
(squares for $T=10^3$, left trangles for $T=10^4$ and crosses for $T=10^5$). 
}
\label{nnnn}
\end{center}
\end{figure}

\begin{figure}[t]
\begin{center}
\vskip 1truecm
\includegraphics[scale=0.4, angle=0]{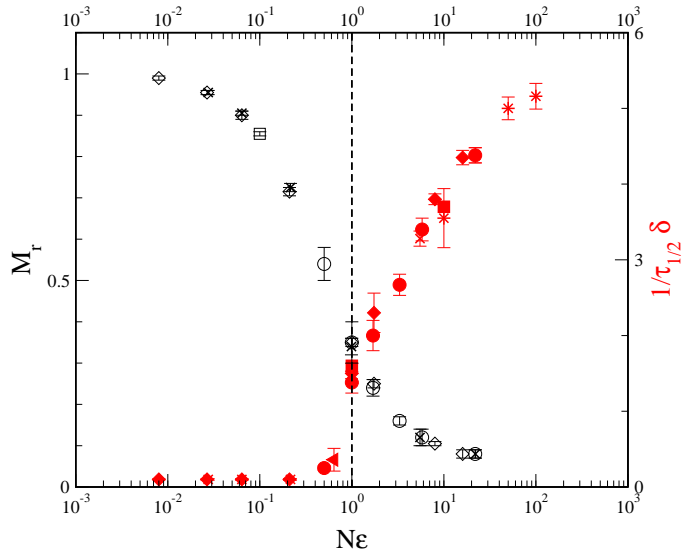}
\caption{Left axis : remnant magnetization $M_r$ {\it vs} 
the parameter $\epsilon N$  indicated with open symbols.
Rigth axis : inverse demagnetization time $\tau_{1/2}$
rescaled by the concentration $\delta$,
{\it vs} the parameter $\epsilon N$ indicated with full symbols.
List of symbols: 
circles ($\epsilon = 0.1, \delta =10^{-3} $), lozanges ($N=80, \delta=0.01$), 
crosses ($N=220, \delta=10^{-3}$), squares ($N=100, R=4$),
asterisks ($N=500, R=20$), left triangles ($\epsilon = 0.01, \delta =0.01 $).
Initially we choose $S_i^z(0)=1$, $i=1,\ldots,N$. In this Figure
an ensemble of $100$ different configurations has been considered.
Each member of the ensemble has been integrated for $10^4$ 
dimensionless time units.
}
\label{se}
\end{center}
\end{figure}

As we have seen in the previous Section, the system dynamics  can be 
described by the parameter, $\epsilon N$, characterizing 
two different dynamical phases, and  describing how much one--dimensional 
a system is. In this Section we will show
that $\epsilon N$ is also a  good (and intensive) 
scaling parameter for the macroscopic properties 
of the quasi 1--D system. 

In order to prove numerically such  argument we need
to find two physical observables that can describe the paramagnetic
and the ferromagnetic phase and to study their  dependence
on the parameter $\epsilon N$.
To this end, let us introduce, on the paramagnetic side, the 
 demagnetization time $\tau_{1/2} $, defined as  the  time at which
the average magnetization decay to one half of its initial value: 
$$\langle M_z ( \tau_{1/2} ) \rangle = (1/2)\langle M_z(0) \rangle. $$ 
In the same way, on the ferromagnetic side,  we introduce 
the ``remnant magnetization'' $M_r$  as the magnetization
left when $\tau \to \infty$,
$$
M_r = \lim_{\tau\to\infty} \langle M_z (\tau) \rangle.
$$
Let us stress that both quantities are physically sound, in the sense
that they are directly and easily measurable.

While it is clear that, 
even if both quantities can be defined only in their
respective phases, they can give useful information when extended
to the other phases. For instance, in the paramagnetic phase
the average magnetization will  depend strongly on the simulation
time ($T$), while in the ferromagnetic phase the demagnetization
time is typically infinite.\footnote{In this case we underestimate its value
putting 
 the maximum dimensionless simulation time, which is $10^4$.}
The dependence of the  average magnetization at fixed time $T$  
can give important information on the ferromagnetic-paramagnetic
transition. Indeed we can expect a weak dependence of the
 average magnetization on the simulation time $T$, 
in the ferromagnetic region $\epsilon N < 1$,
since the presence of quasi-constant of motions freezes the magnetization,
while, in the paramagnetic phase, the average magnetization 
at time $T$,  $<M_z(T)>$, 
 goes to zero as the time grows.
This fact is clearly shown in
Fig.~(\ref{nnnn}), where the average  magnetization $< M_z(T)>$
is plotted {\it vs}  $\epsilon N$
for different  times $T$. As we can see from Fig.~(\ref{nnnn}),
as we increase the  time $T$  the average magnetization remains
almost constant in the ferromagnetic side ($\epsilon N < 1$),
while it goes to zero in the paramagnetic side  ($\epsilon N > 1$),
thus demonstrating a clear signature of the dimensional transition
discussed above.

Since $\epsilon N = \delta R^3$, one can essentially
consider different ways to approach the critical point
 $\epsilon N = \delta R^3 \simeq 1$,
keeping fixed one of the four quantities $\epsilon, N, \delta, R$ and varying
correspondingly the others. This is exactly what we did
in Fig.~\ref{se}, where we show, on the same plot,
 the remnant magnetization  $M_r$ (open symbols
and vertical
axis to the left), and the inverse demagnetization time,  $\tau_{1/2}\delta$, 
rescaled by the concentration
$\delta$  ( vertical axis to the right, full symbols)  both as a function
of the parameter $\epsilon N$. 
Reserving later on the discussion about the time-rescaling with $\delta$, 
let us observe two relevant features.
The first one is the presence of a change of curvature of both curves on approaching
the critical border $\epsilon N = \delta R^3 \simeq 1$.
The second is the scaling of all points in the two curves 
(one for $\tau_{1/2}\delta$, the other for the remnant magnetization $M_r$.)

As for the rescaling of the time, let us observe that,
due to the particular quasi 1--D  geometry, 
and to the low concentration $\delta \ll 1 $, closest  dipoles
give the  major contribution to the energy. For instance,
the configuration with all spins aligned along the $z$--axis will have 
an energy,
\begin{equation}
\label{appp1}
E^\prime \propto \sum_{\langle i,j \rangle} \frac{1}{|r_{ij}|^3},
\end{equation}
where the sum is taken over  $N$ couples $\langle i,j \rangle$
of neighbor dipoles. In other words $E^\prime \sim  N/\langle d\rangle^3 
\sim N \delta$, where
$\langle d\rangle $ is the average distance between two dipoles.

\begin{figure}[t]
\begin{center}
\vskip 1truecm
\includegraphics[scale=0.4,angle=0]{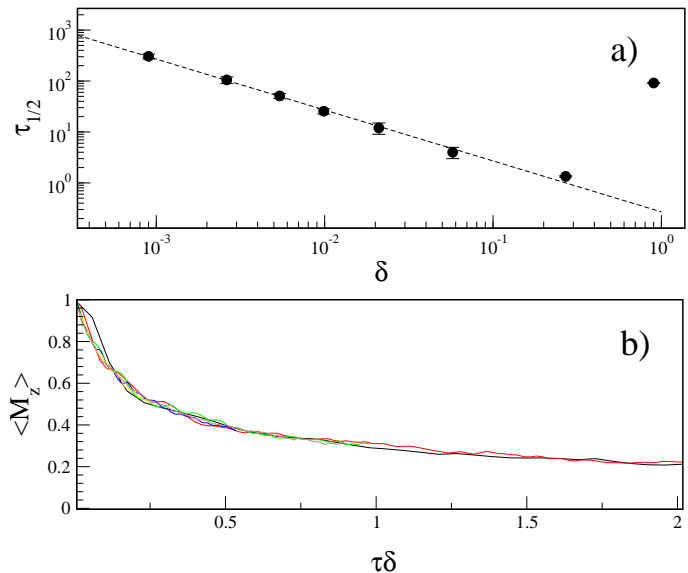}
\caption{a) Dependence of the average  demagnetization
time $\tau_{1/2} $
as a function of concentration $\delta$, for systems with $\epsilon=0.1$
and $N=72$. The average has been taken over an ensemble of $100$
different samples. Initially all samples have all spins aligned along
the $z$--axis : $S_i^z(0)=1, $ \quad $i=1,\ldots N $.
Dashed line represents $\tau_{1/2}  \propto 1/\delta$.
b) Average magnetization $\langle M_z (\tau)\rangle$ as a function
of the rescaled time $\tau \delta$ for different concentration $\delta$
and fixed $\epsilon=0.1$, and $N=72$ as in a).
}
\label{sei}
\end{center}
\end{figure}

On the other hand  the Landau-Lifshitz-Gilbert equations 
of motion are invariant under
a simultaneous scaling of time and energy $\tau' = \tau/\delta$ and 
$E' = E\delta $ so that we will expect  $\tau \propto 1/\delta$.
This simple relation has been verified considering a system  with the
same aspect ratio $\epsilon$, and the same number of particles $N$ 
(so to have the same value of  $\epsilon N$) and 
changing the concentration $\delta$ over 3 orders of magnitude.
Results are presented in Fig.~\ref{sei}a, where $ \tau_{1/2}$
has been shown {\it vs} $\delta$.
To guide the eye a dashed line indicating the inverse
proportionality  has been superimposed. As one can see, looking at the last
point to the right side of Fig.~\ref{sei}a, this relation does not hold true for
concentrations $\delta \simeq 1 $, where the nearest neighbor approximation (\ref{appp1})
fails.

The ``scale invariance'' is even more evident if the average magnetization
is considered as a function of the rescaled time $\tau \delta$, shown in 
Fig.~\ref{sei}b, for the same cases belonging to  the straight line 
shown  in Fig.~\ref{sei}a).

At last, we investigate the system behavior on approaching the large $N$ limit.
First of all let us observe that the scaling variable $\epsilon N = R N/L $ is well
defined in the large $N$ limit  
$N,L \to \infty$, $N/L = const$ and $R$ fixed.\footnote{We thank the Referee for 
this remark.}

\begin{figure}[t]
\begin{center}
\vskip 1truecm
\includegraphics[scale=0.4, angle=0]{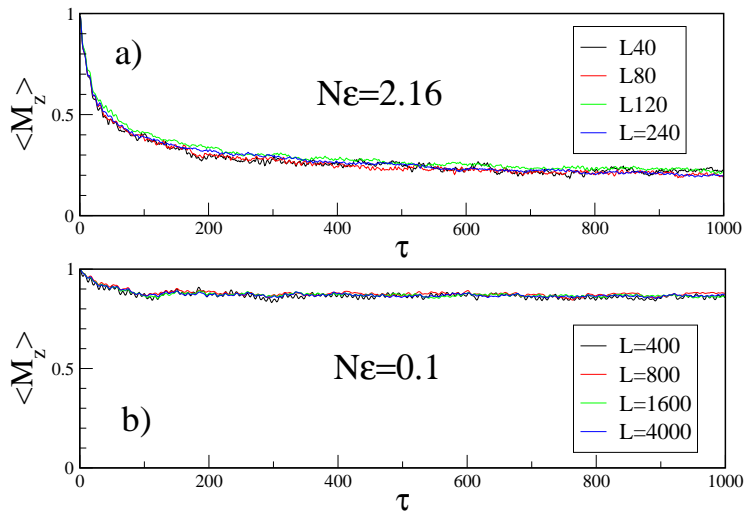}
\caption{Average  magnetization $\langle M_z\rangle$ {\it vs } the dimensionless 
time $\tau$, for different
sample lengths $L$, as indicated in the legend,
 and different number of spins $N$, at fixed density ($N/L=0.36$)
for the paramagnetic phase $N\epsilon =2.16$  (a),
and for the ferromagnetic one $\epsilon N = 0.1$ (b).
In a) is  $R=6$, $\delta=0.01$, while in b) is  $R=4$ 
and $\delta=0.0015625$.
Initially we choose $S_i^z(0)=1$, $i=1,\ldots,N$.
An ensemble of $100$ different configurations has been considered.
}
\label{set}
\end{center}
\end{figure}

In order to do that we take into account different
systems with fixed concentration $\delta$ and radius $R$
and increasing lenght $L$ and number of particles $N$:  both in the 
ferromagnetic phase $\epsilon N <1 $ ( Fig.~\ref{set}a),  
and in the paramagnetic 
one $\epsilon N > 1$ (Fig.~\ref{set}b) , the average
magnetization is independent on the number of particles $N$.

\section{Conclusions}

In this paper the microcanonical dynamics of a system of random dipoles, 
interacting with a pure dipole-dipole interaction has been considered. 
We have shown 
that a dimensional ``phase'' transition, correspondent to a transition 
from regular (ferromagnetic) to stochastic (paramagnetic) regime occurs,
in the microcanonical ensemble, for low concentration $\delta$.
Such transition is  characterized from the dynamical
point of of view by a different   behavior of the fluctuations of the
average magnetization, and from the physical point of view respectively by 
a zero remnant magnetization, 
$ M_r = \lim_{\tau\to\infty} \langle M_z(\tau)\rangle$, and finite decay rates, 
$\propto 1/\tau_{1/2}\delta $ (Paramagnetic Phase)  
or zero decay rates and finite remnant magnetization (Ferromagnetic Phase).
We showed that this dimensional transition
occurs when the intensive parameter $\epsilon N =1$,
where $\epsilon$ is the aspect ratio 
and $N$ is the number of dipoles.
For instance in an experimental situation if we have a
non magnetic substrate with $R=1.6\  nm$ and $L=1.6\  \mu m$,
with lattice size  $\sim 4 0.4 \ nm $, we expect a dimensional 
transition for $\delta= 0.15 \%$.
We also conjectured that in correspondence to this transition,
the ground state changes from ferromagnetic to 
antiferromagnetic.

In the future we would like to investigate dilute dipole 
systems in the canonical ensemble, that
is letting the system be in contact with a thermal bath.
Our analysis in the microcanonical ensemble
indicated that the behavior of very dilute dipoles
in a needle geometry is very similar to a $1$--D 
arrays of dipoles. In the $1$--D case dipole interaction induces
a ferromagnetic ground state, and, due to its anisotropy,
to a breaking of ergodicity \cite{jsp}.
As shown in Ref.~\cite{spada}, the ergodicity breaking 
threshold can induce very large demagnetization  times
thus producing ferromagnetic behavior in finite samples. 
Thus, even if one would expect  that invariant
tori will be destroyed under a suitable
thermal perturbation, the question on the 
demagnetization  times in presence of temperature and 
on the relevance of the ergodicity breaking is still open. 

The ergodicity breaking found
in Ref.~\cite{jsp} considers the total magnetization as an 
order parameter. On the other hand, different  order parameters can be defined 
in dipole systems, 
depending on the ground state configuration, for instance
an anti-ferromagnetic order parameter or a spin glass order
parameter.   Therefore, it would be interesting  
to investigate the existence of an ergodicity breaking
 energy threshold with respect to different order parameters.
 
In conclusion dipole-dipole interacting spin systems offer a  realistic
playground to analyze many properties of magnetic systems
which challenge our comprehension.
  
\section*{Acknowledgments}
We acknowledge useful discussions with S. Ruffo and R. Trasarti-Battistoni.

\end{document}